\documentclass[usenames,dvipsnames,svgnames,table]{aastex7}

\usepackage{graphicx}
\usepackage{mathtools}

\shorttitle{Origins of Carbon Dust at z$\sim$6.7}
\shortauthors{Nanni et al.}

\begin{document}

\title{Origins of Carbon Dust in a JWST-Observed Primeval Galaxy at $z\sim$6.7}

\author[0000-0001-2345-6789]{Ambra Nanni}
\affiliation{National Centre for Nuclear Research, ul. Pasteura 7, 02-093 Warsaw, Poland}
\affiliation{INAF - Osservatorio astronomico d'Abruzzo, Via Maggini SNC, 64100, Teramo, Italy}
\email{ambra.nanni@ncbj.gov.pl}
\author{Michael Romano}
\affiliation{Max-Planck-Institut für Radioastronomie, Auf dem Hügel 69, 53121 Bonn, Germany}
\affiliation{National Centre for Nuclear Research, ul. Pasteura 7, 02-093 Warsaw, Poland}
\affiliation{INAF – Osservatorio Astronomico di Padova, Vicolo dell’Osservatorio 5, I-35122 Padova, Italy}
\email{}
\author{Darko Donevski}
\affiliation{National Centre for Nuclear Research, ul. Pasteura 7, 02-093 Warsaw, Poland}
\affiliation{SISSA, Via Bonomea 265, 34136 Trieste, Italy}
\email{}
\author{Joris Witstok}
\affiliation{Cosmic Dawn Center (DAWN), Copenhagen, Denmark}
\affiliation{Niels Bohr Institute, University of Copenhagen, Jagtvej 128, DK-2200,
Copenhagen, Denmark}
\email{}
\author{Irene Shivaei}
\affiliation{Centro de Astrobiología (CAB), CSIC-INTA, Carretera de Ajalvir km 4, Torrejón de Ardoz 28850, Madrid, Spain}
\email{}
\author{Michel Fioc}
\affiliation{Sorbonne Université, CNRS, UMR 7095, Institut d’astrophysique de Paris, 98 bis boulevard Arago, 75014 Paris, France}
\email{}
\author{Prasad Sawant}
\affiliation{National Centre for Nuclear Research, ul. Pasteura 7, 02-093 Warsaw, Poland}
\email{}
%% Add other authors similarly, e.g.:
%% \author[0000-0002-xxxx-yyyy]{Your Coauthor}
%% \affiliation{Some Institution, Street, Zip City, Country}

%%%%%%%%%%%%%%%%%%%%%%%%%%%%%%%%%%%%%%%%%%%%%%%%%%%%
%%                     ABSTRACT
%%%%%%%%%%%%%%%%%%%%%%%%%%%%%%%%%%%%%%%%%%%%%%%%%%%%
\begin{abstract}
JADES-GS-z6-0, a high-redshift galaxy ($z \sim 6.7$) recently observed as part of the James Webb Space Telescope (JWST) Advanced Deep Extragalactic Survey (JADES), exhibits a distinct bump in its rest-frame ultraviolet (UV) spectrum indicative of a large quantity of hydrocarbon grains, a sign of rapid metal and dust enrichment in its interstellar medium (ISM). This galaxy serves as an ideal case for examining rapid dust formation processes in the early universe.

We investigated diverse dust production channels from a possible maximal formation redshift of $z_{\rm form} \approx 17$, enabling dust contributions from asymptotic giant branch (AGB) stars over the longest possible timescale. Our model simultaneously reproduces key spectral features of JADES-GS-z6-0 such as its Balmer decrement, UV slope, and UV bump. The match is obtained by adopting a star-formation history in which a burst at $\sim 600$~Myr accounts for approximately 30\% of the galaxy's final stellar mass. 

Our findings indicate two pathways for the formation of hydrocarbon grains, such as polycyclic aromatic hydrocarbons (PAHs): (1) efficient dust accretion within the ISM, necessitating a low depletion of metals into dust grains from Type II supernovae ($\approx 10$\%), or (2) dust production predominantly by Type II supernovae, requiring a high depletion fraction ($\approx 73$\%) without dust accretion. We further demonstrate that PAHs are unlikely to originate solely from AGB stars or from shattering of large grains in the ISM. The evolution of the UV slope with redshift points to a complex and bursty star formation history for galaxies observed by JADES.
 \end{abstract}

%% Uncomment keywords if desired
% \keywords{galaxies: high-redshift --- galaxies: ISM --- dust, extinction --- JWST}

%%%%%%%%%%%%%%%%%%%%%%%%%%%%%%%%%%%%%%%%%%%%%%%%%%%%
%%                   MAIN TEXT
%%%%%%%%%%%%%%%%%%%%%%%%%%%%%%%%%%%%%%%%%%%%%%%%%%%%
\section{Introduction}
Cosmic dust provides key information on the evolution of the interstellar medium (ISM) of galaxies. The James Webb Space Telescope (JWST) has emerged as a transformative tool, providing unparalleled capabilities for investigating the role of cosmic dust in the context of galaxy evolution. %Recent findings from the JWST Advanced Deep Extragalactic Survey \citep[JADES,][]{Eisenstein23} have unveiled the presence of the 2175 $\AA$ dust feature (UV bump) in galaxies reaching redshifts up to $z\sim7$ which suggests the existence of small carbonaceous grains, namely polycyclic aromatic hydrocarbons (PAHs), in these early epochs \citep{Witstok23}. 
Recent JWST observations have unveiled a prominent absorption feature, referred to as the ``UV bump'', around 2175 $\mathrm{\AA}$ (rest-frame), in primeval galaxies up to redshifts of $z\sim8$ \citep{Witstok23, markov24}. This discovery points to the presence in the early universe of small carbonaceous grains, possibly polycyclic aromatic hydrocarbons (PAHs; \citet{yang25}), which are thought to produce this distinctive spectral feature \citep{lin23}.

Of particular interest is the galaxy JADES-GS+53.15138-27.81917 (hereafter JADES-GS-z6-0), observed as part of JWST Advanced Deep Extragalactic Survey \citep[JADES;][]{Eisenstein23}. JADES-GS-z6-0 is identified at $z\sim 6.7$, corresponding to a cosmic age of only 800 million years (Myr) post-Big Bang. The galaxy exhibits an unusually strong UV bump, comparable to those observed in the Milky Way and the Large Magellanic Cloud, which may indicate rapid carbon enrichment potentially driven by Type II Supernovae (SNII) and/or Wolf-Rayet stars \citep{Mathis94}. Although no definitive carrier has been identified, \citet{Li24} examined whether small graphite grains could explain the extinction bump observed in JADES-GS-z6-0 and concluded that the resulting feature would not match the observed one. In contrast, \citet{Lin25} showed that a mixture of different PAH molecules closely reproduces the bump.

In this work, we aim to investigate the origin of the UV bump observed in JADES-GS-z6-0 with a model of chemical evolution of galaxies coupled with radiative transfer modelling of the stellar radiation through dust. Our primary goal is to simultaneously reproduce the multiple spectral features detected in the NIRSpec data for JADES-GS-z6-0 and to investigate the physical processes responsible for PAH formation. Specifically, we examine whether abundant carbonaceous grains could originate from SNII ejecta or from low- and intermediate-mass stars (with masses $\lesssim 8-10$~M$_\odot$) evolving through the asymptotic giant branch (AGB) phase, as well as the potential contribution of grain growth within the ISM.
Throughout the paper, we assumed a flat $\Lambda$CDM cosmology based on the results of the Planck collaboration, i.e. H$_0=67.4$~km~s$^{-1}$~Mpc$^{-1}$, $\Omega_{\rm m} = 0.315$ \citep{Planck20} and a Chabrier initial mass function (IMF) \citep{Chabrier03} with lower and upper limits of 0.08 and 120~$M_\odot$ which maximizes the number of low- and intermediate-mass stars that evolve through the AGB phase, thereby providing an upper limit on dust production by AGB stars in JADES-GS-z6-0. Data were retrieved from the MAST archive DOI: 10.17909/rm0h-xz71.

%--------------------------------------------------------------------
\section{Method}

\subsection{Metal and dust evolution in the ISM}
We track the dust evolution in the ISM of JADES-GS-z6-0 by following the model introduced by \citet{Dwek98} to which we refer for all the details. We use for this scope the code \textsc{P{\'e}gase.3} \citep{Fioc19} to consistently compute the metal and dust evolution. This model accounts for metal and dust injection from stellar sources, dust destruction by supernova (SN) shocks, and dust growth through grain accretion in the ISM. For this study, we adopt a closed-box model. This choice is motivated by the observed similarities between JADES-GS-z6-0 and local dwarf galaxies \citep[DGS;][]{Madden13}, for which \citet{Romano23} found mass-loading factors (i.e., the mass outflow rate of the galaxy divided by its SFR) for the atomic gas component typically ranging between 0 and 1. 

We computed the dust evolution of silicates and carbonaceous grains, including polycyclic aromatic hydrocarbons (PAHs), which are assumed to constitute a fraction of the total dust mass. Specifically, we set the PAHs abundance to be proportional to the metallicity following the empirical relation by \citet{Galliano21}\footnote{We assumed that PAHs are the carriers of the UV bump and, following the THEMIS calibration, adopt qPAH = 0.45 $\times$ qAF, where qAF is the fraction of hydrogenated amorphous carbon grains responsible for the mid-IR emission \citep[see also][for details.]{Galliano21}}:
\begin{equation}\label{Eq:mPAH}
    m_{\rm PAHs}=0.45 \times 10^{-9.001 + 0.9486\times(12+\log \frac{O_{\rm ISM}}{H_{\rm ISM}})}M_{\rm dust},
\end{equation}
where $M_{\rm dust}$ is the total dust mass given by silicates plus carbonaceous species. We vary the value of $m_{\rm PAHs}$ between a minimum and maximum value that includes 95\% of the sources in \citet{Galliano21}. This is obtained by adding to the exponent of Eq.~\ref{Eq:mPAH} values between $[-0.36, +0.20]$.

We defined the depletion fraction as the proportion of carbon atoms incorporated into carbon grains and the fraction of Mg, Si, S, Ca, and Fe atoms incorporated into silicate grains. For silicate formation, we assume an oxygen atom ratio $n_{\rm O}/n_i = 1$ for each element $i$ involved. For supernovae (for both Type II and Ia), we apply a variable depletion fraction ($f_{\rm SN}$), while for AGB stars, we adopt a fixed value of $50\%$ for both carbonaceous and silicate dust, based on circumstellar dust formation studies \citep[e.g.,][]{Ventura12, Nanni13}. Finally, while the \textsc{P{\'e}gase.3} code does not include metal enrichment from Population III stars, prior studies suggest that this contribution is likely negligible \citep[e.g.,][]{Nanni20}.

We introduced a few modifications with respect to the standard prescriptions in \textsc{P{\'e}gase.3}, in particular for the dust destruction time-scales due to SNe shocks and for dust growth due to accretion in the ISM. Namely, for the mass of gas swept up by each SN event, we adopted the relation from \citet{Asano13}:
\begin{equation}
    M_{\rm swept} = 1535 ~n_{\mathrm{SN}}^{^{-0.202}}[Z/Z_{\odot} + 0.039]^{-0.289},
\end{equation}
where $n_{\mathrm{SN}}$ = 1.0 $\rm cm^{-3}$ is the gas density around SNe of type Ia and II, and where the solar metallicity is assumed to be $Z_\odot=0.014$ \citep{Asplund21}.

Dust accretion in the ISM is computed in \textsc{P{\'e}gase.3} by defining a certain accretion timescale that we calculated as \citet{Asano13}:
\begin{equation}
    \tau_{\rm accr}= 2.0 \times 10^7 \Big(\frac{Z}{0.02}\Big)^{-1} [{\mathrm yr}].
\end{equation}
%where $\tau_{\rm accr, 0}= 2.8\times10^5$~yr. We note that this value differs from the one by \citet{Asano13} due to different value of the solar metallicity (0.014 rather than 0.02) in their Eq. 20.

Finally, in order to match the final predicted gas-phase metallicity with the one obtained from nebular measurements in \citet{Witstok23} ($Z_{\rm neb}\sim 0.2~Z_{\odot}$), we select the initial mass of gas to be $M_{\rm gas, ini}\sim 12$ times the final mass of stars.

\subsection{Star formation history} 
We considered the general functional form delayed plus burst of the star formation history (SFH).
A simple delayed SFH is defined as
\begin{equation}\label{Eq:SFH_del}
        \mathrm{SFR}_{\rm del}(t)=k\frac{t}{\tau_{\rm del}^2}\exp\Big(-\frac{t}{\tau_{\rm del}}\Big),
\end{equation}
where $t$ is the age of the stellar population assumed to be formed at $t=0$~Myr, $k$ is a normalisation factor, and $\tau_{\rm del}$ is the time at which the SFR peaks. 
Following the definition in Eq.~\ref{Eq:SFH_del}, the functional form of the delayed plus burst SFH reads:
\begin{equation}\label{Eq:SFH_del_b}
\centering
\mathrm{SFR}_{\rm del+burst} (t)=
\begin{cases}
     \mathrm{SFR}_{\rm del}(t){\rm, \, if \, \, t<t_{\rm burst}} \\
     \mathrm{SFR}_{\rm del}(t)+ k_{\rm burst }\exp\Big(-\frac{t}{\tau_{\rm burst}}\Big) {\rm, \, \, if \, t\geq t_{\rm burst}},
\end{cases}
\end{equation}
where $t_{\rm burst}$ is the age of the galaxy at which the burst starts, $\tau_{\rm burst}$ is the e-folding time of the young stellar population component formed during the burst episode, and $k_{\rm burst}$ is determined by the fraction, $f_{\rm burst}$, of stars formed in the burst episode with respect to the total as:
\begin{equation}
    k_{\rm burst}=\frac{f_{\rm burst}}{1-f_{\rm burst}}\frac{\int^{\theta}_0 \mathrm{SFR}_{\rm del}(t)\,dt}{\int^{\rm \theta}_{t_{\rm burst}} \exp(-\frac{t}{\tau_{\rm burst}})\,dt},
    \end{equation}
where $\theta$ is the age of the galaxy.
We assumed an upper limit for the formation redshift of the galaxy of $z_{\rm form}\sim 17$, which implies an age of $\approx 600$~Myr. These assumptions allow AGB stars to maximize their time to produce dust.

\subsection{Stellar population synthesis model and yields}
We constructed composite stellar populations consistently with the metallicity evolution using \textsc{P{\'e}gase.3}, which employs the ``Padova'' stellar evolutionary tracks from \citet{Bressan93}, \citet{Fagotto94a, Fagotto94b, Fagotto94c} and \citet{Girardi96}. Stellar spectra are sourced from the BaSeL library v3.1 \citep{Kurucz79, Westera02} for stars with effective temperatures $T_{\rm eff}<50000$~K, while spectra for hotter stars are obtained from \citet{Rauch03}.
We adopted metal yields from \citet{Portinari98} and \citet{Marigo01}, which are consistent with our chosen stellar tracks.\footnote{The lower limit for the metal yields in \citet{Marigo01} is $Z=0.004$. We tested extrapolated metallicity yields down to $Z=0.0001$ \citep{Fioc19}. The difference in the final metallicity value is within $1$\%. We therefore selected $Z=0.004$ both for the metal yields and tracks for our analysis.}
Moreover, we tested different combinations of Type II SNe and AGB yields by using the \textsc{omega} code \citep{Cote17} in order to assess which set provides the maximal carbon return from AGB stars. We selected all the possible combinations provided in Table~\ref{Table:models}. We found that, for final metallicity in agreement with the one derived from observations \citep{Witstok23}, the combination with \citet{Portinari98} for Type II SNe and \citet{Marigo01} for AGB stars maximizes the carbon return from AGB stars, therefore providing an upper limit for carbon formation.\footnote{The metallicity values derived with \textsc{omega} and \textsc{P{\'e}gase.3} differ by about 20\% for the same selected yields, which is within the observational uncertainty.}

\subsection{Stellar and dust distributions and radiative transfer}
We performed the radiative transfer calculations of stellar radiation through dust using the \textsc{radmc-3d} Version~2.0 \citep{Dullemond13}. To achieve this goal, we computed the dust-free spectra as a function of age with \textsc{P{\'e}gase.3}. We assumed as input in \textsc{radmc-3d} a smooth stellar source distribution, using the dust-free integrated stellar population.

The stellar density distribution was assumed to be spherical and was derived from the surface brightness profile of stars observed with JWST NIRCam in the F356W filter, covering the rest-frame optical continuum at $\sim$4700~$\mathrm{\AA}$. To extract the surface brightness profile, we used the \textsc{galfit} package and we assumed a modified King distribution \citep{Elson99}, suitable for spheroidal dwarf galaxies \citep{Burkert15}:
\begin{equation}\label{king}
\begin{multlined}
    \Sigma = \Sigma_0 \Big[1-\frac{1}{(1+(R_{\mathrm t}/R_{\mathrm c})^2)^{1/\alpha}}\Big]^{-\alpha} \times \\
    \Big[\frac{1}{(1+(r/R_{\mathrm c})^2)^{1/\alpha}}- \frac{1}{(1+(R_{\mathrm t}/R_{\mathrm c})^2)^{1/\alpha}}\Big]^{\alpha},
\end{multlined}
\end{equation}
where $\Sigma_0\approx 9 \times 10^{-26}$~mag$/$arcsec$^2$.
The point-spread function (PSF) was generated with WebbPSF \citep{Perrin12} in the NIRCam F356W filter\footnote{\url{https://stsci.app.box.com/v/jwst-simulated-psf-library}}.
We determined the core radius, $R_{\rm c}=630$~pc, the truncation radius, $R_{\rm t}=1200$~pc, and the value of $\alpha$ for the stellar distribution, $\alpha_{\rm stars}=3$.

We assumed the dust density distribution for each dust species (silicates, graphite and PAHs) to have an analogue empirical King profile as the stellar distribution but defined with $\alpha_{\rm dust}=C\times \alpha_{\rm stars}$, where $C=0.8$. The value of $C$ was derived from the fit of UV and FIR profiles in local dwarf galaxies that trace the stellar and dust components, respectively \citep{Romano24}, and was based on the assumption that local, low-metallicity galaxies are analogues of high-redshift star-forming sources (e.g., \citealt{Flores21, Shivaei22}).

The absorption coefficients for the different dust species were integrated on the grain size distributions adopted in \citet{WD01} and \citet{DL07}. For this calculation we integrated the absorption coefficients as a function of the grain size from \citet{Draine84} and \citet{Laor93} for astronomical silicate and graphite grains, and from \citet{Li01} for PAHs. We adopted the set of absorption coefficient for neutral PAHs that are similar to ionised ones in the UV-optical bands. These absorption coefficients are provided in the wavelength range 0.001--1000~$\mu$m, which covers the entire spectrum of the computed stellar population\footnote{Absorption coefficients were downloaded from \url{https://www.astro.princeton.edu/~draine/dust/dust.diel.html}}. We include in the calculation isotropic scattering and absorption and we adopted the modified random walk approximation.
The final spectra are then reprocessed for the transmissivity of the intergalactic medium at the redshift of the observed galaxy \citep{Inoue14}.

\subsection{Nebular lines}
We computed the nebular emission for the dust-free case by means of \textsc{P{\'e}gase.3} whose calculations are based on pre-computed grids of models from \textsc{cloudy} \citep{Ferland17}.
The gas electron density is assumed to be constant throughout the cloud $n_e=100$~cm$^{-3}$. We assumed a variable fraction of the Lyman continuum photons absorbed by the gas \citep[$f_{\rm LyC, gas}$; ][]{Fioc19}. The remaining fraction of Lyman continuum photons are either absorbed by dust or escape the galaxy. The calculations take into account the variation of metallicity of the ISM. We finally estimated the attenuated lines based on the continuum attenuation with respect to the dust-free case.

\subsection{Characterising JADES-GS-z6-0}
We constructed a model grid by discretizing six key parameters: the burst fraction ($f_{\rm burst}$), the nebular gas fraction ($f_{\rm neb}$), the supernova feedback parameter ($f_{\rm SN}$), the stellar mass normalization ($M_{\rm norm}$), the age, and $k$ in Eq.~\ref{Eq:SFH_del_b}. We built grids of spectra by varying the model parameters as summarized in Table~\ref{Table:models}. We took into consideration only models with $H\alpha/H\beta>3.6$. For each combination in the parameter grid, we computed model observables (i.e., nebular line fluxes and dust-attenuated continuum). Because our grid is discretely sampled, we employed a ``snapping'' procedure to map any continuous parameter value proposed by the Markov Chain Monte Carlo (MCMC) sampler to the closest available grid point.  We excluded the [OIII] lines from the $\chi^2$ a posteriori estimates because their high O$\lambda$5008/H$\alpha$ ratio was not well reproduced by our preliminary calculations, even assuming the maximum possible ionisation parameter in the diffuse ISM (see Table~\ref{Table:models}). This suggests the presence of a radiation field harder than what our models currently included.

The posterior probability distribution of the six parameters was then explored using MCMC with the \texttt{emcee} package. The likelihood function was defined in terms of the $\chi^2$ value, which is calculated by comparing the observed spectrum (after correcting for intergalactic medium transmission) with the model continuum and by comparing the model’s emission-line fluxes with their observed values.
After running the MCMC, we computed the uncertainties on both the input parameters and the derived quantities by calculating the 16th, 50th (median), and 84th percentiles of their respective posterior distributions.

\begin{table*}
\begin{center}
\caption{Metal yields adopted in the chemical evolution code \textsc{omega}, to select the combination of Type II SN and AGB yields which maximizes the carbon production from AGB stars. All possible combinations have been tested. The input parameters for the calculations of the dust-free spectrum, nebular lines and chemical evolution with \textsc{P{\'e}gase.3} are also provided.}
\label{Table:models}
\begin{tabular}{l l l}
     & Tested yields & \\
    \hline
    Type II SN                      & AGB                          &   \\
  \hline
  \citet{Portinari98}  &   \citet{Marigo01} & \\
  \cite{LC18}, \citet{Prantzos18}   &   \citet{Cristallo15} & \\
  \citet{Ritter18, Fryer12} delayed  &   \citet{Karakas10} & \\
  \citet{Ritter18, Fryer12} rapid & \citet{Ritter18} & \\
  \citet{Kobayashi06} & \\

    \multicolumn{3}{c} {Stellar continuum}\\
    \hline
    Quantity                       & Denomination &  Values/References                             \\

  \hline
  Stellar tracks &   & \citet{Bressan93, Fagotto94a, Fagotto94b, Fagotto94c}\\                       &     & \citet{Girardi96}  \\
  Initial Mass Function &  &\citet{Chabrier03}  \\ 
   e-folding time-scale of the delayed SFH & $\tau$ [Myr] &100 \\ 
  e-folding time-scale of the burst & $\tau_{\rm burst}$ [Myr] &10 \\ 
 Mass fraction of stars formed during the burst & f$_{\rm burst}$  &  9 values in [0.1-0.5] \\ 
 Burst age &   age$_{\rm burst}$   [Myr]     & 594 \\
 Normalisation factor & $k$~[M$_\odot$] & 6 values in [1.4-2.4]\\
 Age of the galaxy                   &      age  [Myr]   & 11 values in [594-604]\\
    \multicolumn{3}{c}{}\\
    \multicolumn{3}{c}{Nebular emission}\\
  \hline
      Quantity                       & Denomination &  Values/Reference                             \\
\hline
  Ionisation parameter in the diffuse ISM &  $\log_{10} U$    &  $-1$  \\
 % Gas metallicity &  Z$_{\rm gas}$ & 0.004 \\
  Electron density & $n_{\rm e}$~[cm$^{-3}$] & 100 \\
  Fraction of Lyman continuum photons absorbed by gas & $f_{\rm LyC, gas}$ & 19 values in [0.05-0.95] \\
  Other & & Default values in \textsc{P{\'e}gase.3}  \\
   \multicolumn{3}{c}{}\\
    \multicolumn{3}{c}{Chemical evolution}\\
  \hline
     Quantity                       & Denomination &  Values/References                             \\
\hline
    Metal yields          &   &\citet{Portinari98, Marigo01}   \\
Final stellar mass &   M$_{\rm star, fin}$ [M$_\odot$]    & 10 values in [$10^8$--$10^9$] \\
   Initial mass of gas &   $M_{\rm gas, ini}$ [M$_{\rm star, fin}$]   & 12\\

  AGB depletion fraction & f$_{\rm AGB}$ & 50\% \\
  Dust growth in the ISM &                &  yes, no   \\
 SNe depletion fraction (with dust growth in the ISM) & f$_{\rm SN}$ & 0.01, and 8 values in [0.025-0.2] \\
 SNe depletion fraction (without dust growth in the ISM) & f$_{\rm SN}$ & 11 values in [0.5-1] \\

\hline

\end{tabular}
\end{center}
\end{table*}

\begin{table*}
\begin{center}
\caption{Derived physical properties for JADES-GS-z6-0 based on the latest MCMC analysis, comparing scenarios \emph{with} and \emph{without} dust growth in the ISM, alongside literature values when available.  All masses are given in solar masses (M$_\odot$).  Uncertainties indicate the 16th and 85th percentiles.}
\label{Table:results}
\begin{tabular}{l c c c}
\hline
Quantity & With growth & Without growth & Literature \\
\hline
\multicolumn{4}{l}{\emph{Stellar population properties}}\\[5pt]
Final stellar mass (M$_*$ [M$_\odot$])  
& $\bigl(7.8^{+1.1}_{-1.8}\bigr)\times10^{8}$  
& $\bigl(7.4^{+1.7}_{-3.0}\bigr)\times10^{8}$  
& $1.0^{+0.3}_{-0.2}\times10^{8}$ \citep{Witstok23} \\[3pt]
Burst fraction ($f_{\rm burst}$) 
& $30^{+2}_{-2}\,\%$ 
& $24^{+3}_{-5}\,\%$
& -- \\[3pt]
LyC photons absorbed by gas ($f_{\rm LyC, gas}$)
& $39^{+3}_{-5}\,\%$ 
& $27^{+8}_{-3}\,\%$  
& -- \\[5pt]
Depletion fraction for SN ($f_{\rm SN}$)
& $10^{+1}_{-1}\,\%$ 
& $73^{+4}_{-19}\,\%$
& - \\[5pt]
Mass-averaged age$^\dagger$ [Myr]
& $293^{+4}_{-0}$ 
& $325^{+13}_{-9}$ 
& $18^{+11}_{-7}$ \citep{Witstok23} \\[5pt]
\multicolumn{4}{l}{\emph{Star formation and attenuation}}\\[5pt]
Star formation rate (SFR [M$_\odot$ yr$^{-1}$]) 
& $16^{+4}_{-4}$ 
& $15^{+7}_{-7}$ 
& $3^{+2}_{-1}$ \citep{Witstok23}\footnote{SFR in Witstok et al.\ (2023) is averaged over 30\,Myr.} \\[3pt]
$H\alpha/H\beta$ 
& $3.71^{+0.03}_{-0.03}$ 
& $3.81^{+0.06}_{-0.15}$ 
& $4.15 \pm 0.72$ \citep{deugenio24} \\[3pt]
UV slope ($\beta$)
& $-2.24^{+0.01}_{-0.02}$
& $-2.15^{+0.05}_{-0.02}$
& $-2.13 \pm 0.18$  \citep{Witstok23} \\[3pt]
Maximum extinction ($A_{\rm max}$ [mag]) 
& $0.33^{+0.00}_{-0.00}$ 
& $0.33^{+0.01}_{-0.03}$ 
& $0.43 \pm 0.07$  \citep{Witstok23} \\[3pt]
Central wavelength ($\lambda_{\rm max}$ [\AA]) 
& $2168^{+0.4}_{-1.1}$ 
& $2169^{+4}_{-1}$ 
& $2263^{+20}_{-24}$  \citep{Witstok23} \\[5pt]
\multicolumn{4}{l}{\emph{Dust properties}}\\[5pt]
Dust depletion for SNe ($f_{\rm SN}$) & $10^{+1}_{-1}$\% & $73^{+4}_{-19}$\%  & - \\
Total dust mass (M$_{\rm dust}$ [M$_\odot$]) 
& $\bigl(6.3^{+0.8}_{-1.6}\bigr)\times 10^{6}$ 
& $\bigl(4.6^{+2.6}_{-1.9}\bigr)\times 10^{6}$ 
& -- \\[3pt]
Carbon dust mass (M$_{\rm carb}$ [M$_\odot$]) 
& $\bigl(2.3^{+0.3}_{-0.6}\bigr)\times 10^{6}$ 
& $\bigl(1.2^{+0.6}_{-0.5}\bigr)\times 10^{6}$ 
& -- \\[3pt]
Silicate dust mass (M$_{\rm sil}$ [M$_\odot$])
& $\bigl(3.8^{+0.5}_{-1.0}\bigr)\times 10^{6}$ 
& $\bigl(3.2^{+1.8}_{-1.3}\bigr)\times 10^{6}$ 
& -- \\[3pt]
PAH dust mass (M$_{\rm PAH}$ [M$_\odot$])
& $\bigl(2.5^{+0.3}_{-0.6}\bigr)\times 10^{5}$ 
& $\bigl(2.2^{+1.1}_{-1.1}\bigr)\times 10^{5}$ 
& -- \\[3pt]
PAH mass fraction ($f_{\rm PAH}$)
& $4.0^{+0.0}_{-0.1}\,\%$ 
& $4.6^{+0.0}_{-0.4}\,\%$ 
& -- \\[3pt]
\hline
\end{tabular}
\end{center}
\end{table*}

\section{Results}\label{Sec:results}
\subsection{Reproducing the spectrum of JADES-GS-z6-0 }
We reproduced most of the properties of the spectrum (UV slope, Balmer decrement and UV bump) by fixing $\tau=100$~Myr and $\tau_{\rm burst}= 10$~Myr, where the burst occurs after 594 Myr. The recent burst of star formation producing a young stellar component is needed to reproduce the steep observed UV slope \citep{Witstok23}.

Given the uncertainties affecting our knowledge of dust accretion in the ISM of galaxies and the controversial results related to its efficiency \citep[e.g.][]{Rouille15, Ceccarelli18,Priestley21}, we considered two scenarios for dust enrichment in the ISM: by including dust growth in the ISM and without any dust growth in the ISM. We neglected the possible contribution to the total dust budget from the winds of massive stars for this calculation. 
The physical parameters of the galaxy are given in Table~\ref{Table:results}. 

For both simulations, we set the initial gas mass to be 12 times the final stellar mass, resulting, without outflow, in a final gas mass of roughly 11 times the final stellar mass. We thus derive dust-to-gas ratios of $\sim 5.4-5.6 \times 10^{-4}$, consistent with values found in other high-redshift galaxy studies \citep[e.g][]{Palla24,Sawant24}.

\subsubsection{Models with dust growth in the ISM}
\begin{figure}
   \centering
   \vspace*{\fill}%
      \includegraphics[scale=0.5]{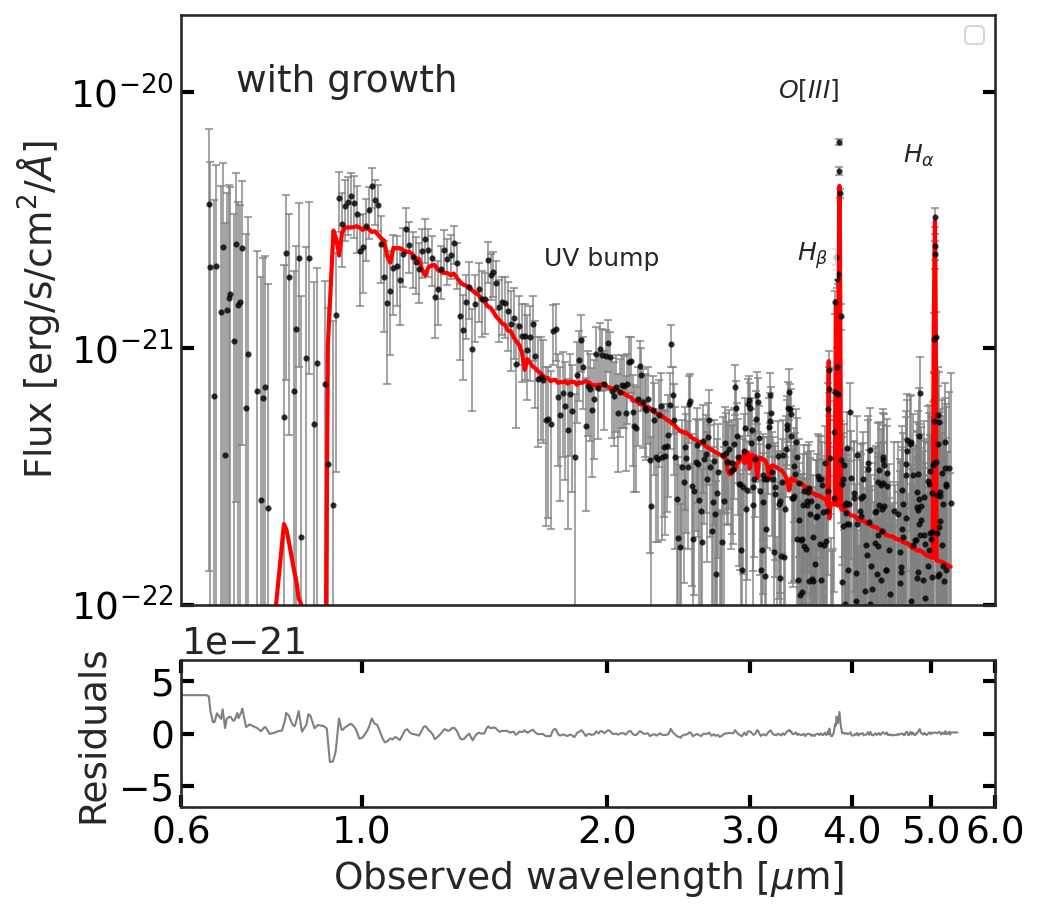}
         \includegraphics[scale=0.5]{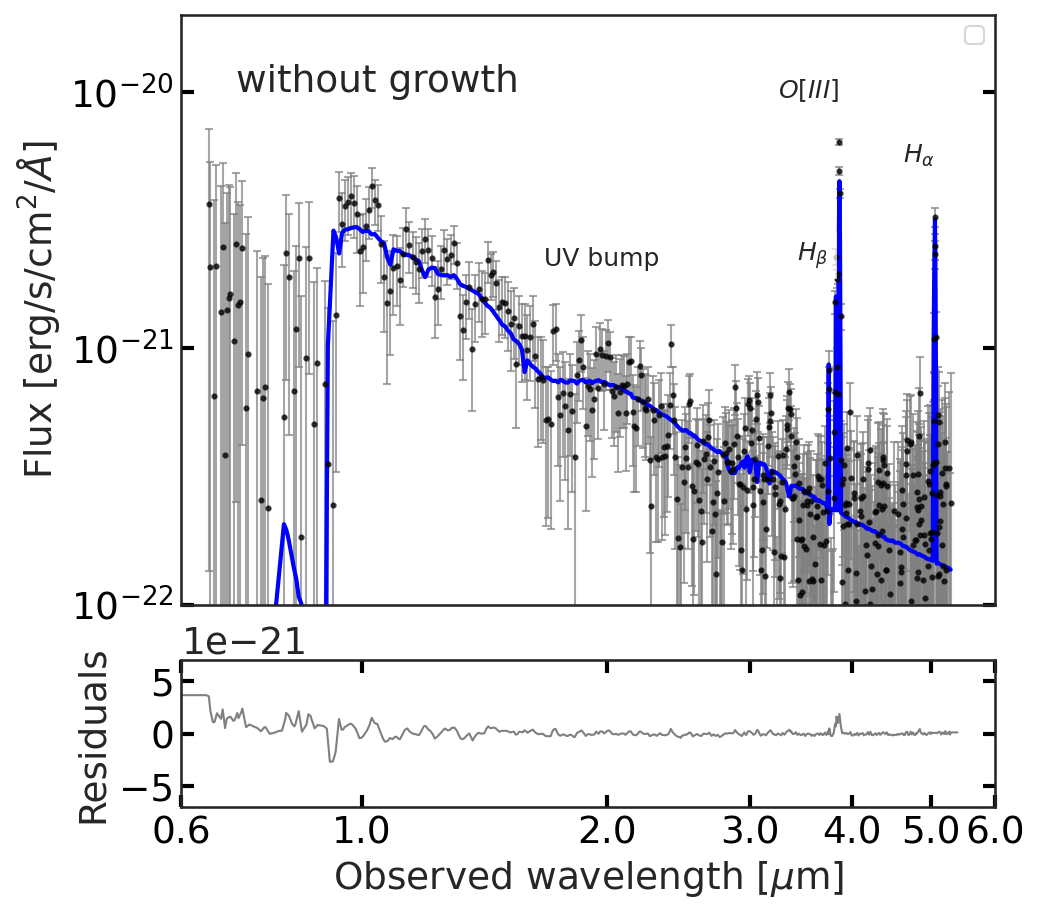}
     \vspace*{\fill}%
   \caption{Comparison between the observed spectrum of JADES-GS-z6-0 (black points with uncertainties) and the best-fit predicted spectrum from self-consistent chemical evolution of the ISM: (left panel) spectrum predicted with $f_{\rm burst}=30$\%, $f_{\rm SN}=10$\%, $M_{\rm *}= 8 \times 10^8$~M$_\odot$ and dust growth in the ISM (red curve); (right panel) spectrum predicted with $f_{\rm burst}=25$\%, $f_{\rm SN}=75$\%, $M_{\rm *}= 9 \times 10^8$~M$_\odot$ and no dust growth in the ISM (blue curve).}
              \label{Fig:spectra}
   \end{figure}

The observed and the best-fitting predicted spectra are shown in Fig.~\ref{Fig:spectra} (left panel) for models with dust growth in the ISM.  
We selected models that reached the final mass of the galaxy for ages $>593$~Myr, during the bursting phase. The continuum, including the UV slope, is well reproduced by our model.
The stellar mass derived in this work is $\sim 7.8$ times larger than that reported by \citet{Witstok23}. We verified that such a difference is likely due to the assumed star formation history (SFH) in the present work that extends to 600 Myr. Another source of difference can be due to the different attenuation laws adopted in \citet{Witstok23} and derived here. Similarly, the mass-averaged age of the stellar population is significantly older than what was derived from the spectral energy distribution (SED) fitting by \citet{Witstok23}. The UV slope and $H\alpha/H\beta$ are reproduced within the uncertainties. We here emphasise that, for the goal of the paper, we adopted a SFH maximizing the contribution of AGB stars, and not due to best-fit approach.

We found from our radiative transfer calculations a flux of 0.6~$\mu$Jy at $\lambda=$1.2 mm, which is consistent with the non-detection by ALMA \citep{Witstok23}.
The abundance of PAHs required to match the observed UV bump is 
$\sim 4\%\;\text{of the total dust mass}$,
which is consistent with the upper limit of PAH abundance observed in local galaxies \citep{Galliano21} at a given metallicity value. We fitted the excess attenuation relative to the continuum of the simulated spectra using a Drude profile \citep{Fitzpatrick86}, where the continuum was modelled as a power law with the UV slope $\beta$ derived from spectral fitting.  
We found a maximum extinction relative to the continuum of 
$A_{\rm max} \sim 0.33$,
which is consistent within $2\sigma$ with the value derived from the observed spectrum by \citet{Witstok23}. This can be either due to the underestimation of the small carbon mass fraction or to a different geometry and/or line of sight with respect to the one assumed.
The central wavelength of the predicted profile is also shifted to lower values than the observed one \citep{Witstok23}. This shift toward shorter wavelengths has already been noticed by \citet{Li24} for small graphite grains.
Hydrogen nebular lines are reproduced within uncertainties \citep{deugenio24}, assuming that 
$f_{\rm LyC, gas} \sim 39\%$. However, oxygen lines are under-predicted by a factor of $\sim 1.3$--$1.4$.
For the best fitting model, we computed the escape fraction of Lyman continuum photons ($f_{\rm LyC, esc}$): we obtained on average $f_{\rm LyC, esc}\sim 2\%$.
\subsubsection{Models without dust growth in the ISM}
The best-fitting predicted and observed spectra are shown in Fig.~\ref{Fig:spectra} (right panel) for models without dust growth in the ISM. The galaxy's estimated physical parameters closely match those found for the scenario where dust growth is included, but for the depletion fraction of SNe, which is higher to reproduce the Balmer decrement in absence of dust growthin the ISM. The results are shown in Table~\ref{Table:models}.
We found that the high fraction of dust condensation required for SNe, $f_{\rm SN}\sim 73$\% in this scenario is in agreement with other findings in the literature for local and high-redshift galaxies \citep{DeLooze20, Burgarella20, Nanni20, Sawant24}.
Hydrogen lines can be reproduced within the uncertainties with $f_{\rm LyC, gas} \sim 27\%$. However, oxygen lines remain under-predicted by about the same factor as in the scenario with dust growth in the ISM. This shortfall could point to an additional ionizing source, such as an AGN, or to a top-heavy IMF. Alternatively, the discrepancy may arise from the models' omission of Wolf-Rayet stars, whose hard ionizing spectra can significantly enhance high-ionization lines such as OIII.
For the best fitting model, we estimated the escape fraction of Lyman continuum photons similarly to the previous case, obtaining $f_{\rm LyC, esc}\sim 2\%$.

\subsection{Carbon dust production}
We first tested whether dust production from AGB stars alone could reproduce the observed attenuation by considering an extreme scenario in which only AGB stars produce dust (i.e.\ neglecting dust by SNe and ignoring dust growth in the ISM). We found that dust from AGB stars only (reduced by destruction by SN shocks) provides a minor contribution to the total dust extinction for the model, whose best-fitting continuum yields $H\alpha/H\beta<3$. This indicates that a fraction of dust must be produced by Type~II~SNe and/or that some fraction of dust has grown in the ISM.

\begin{figure}
   \centering
   \vspace*{\fill}%
      \includegraphics[scale=0.42]{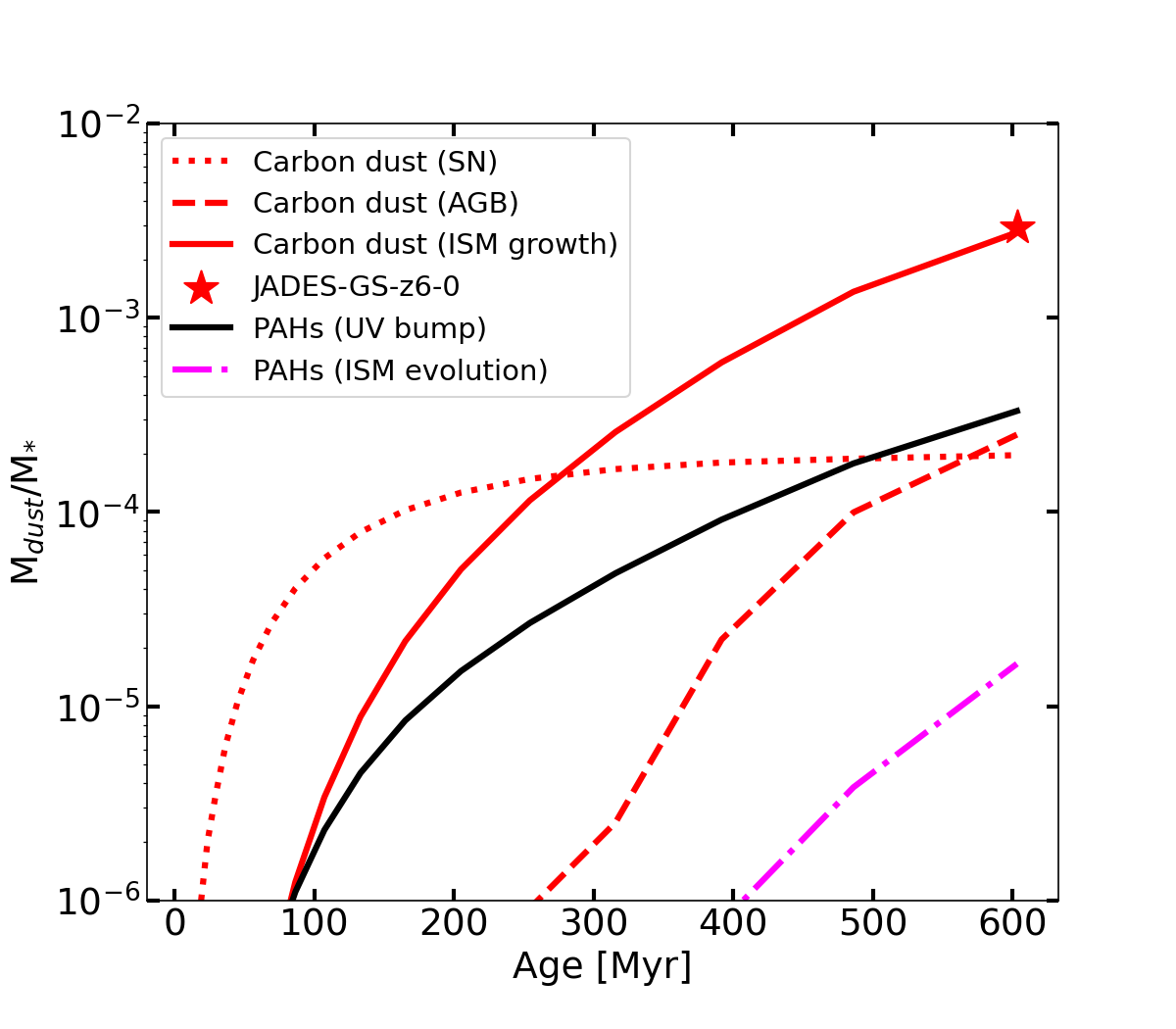}
         \includegraphics[scale=0.42]{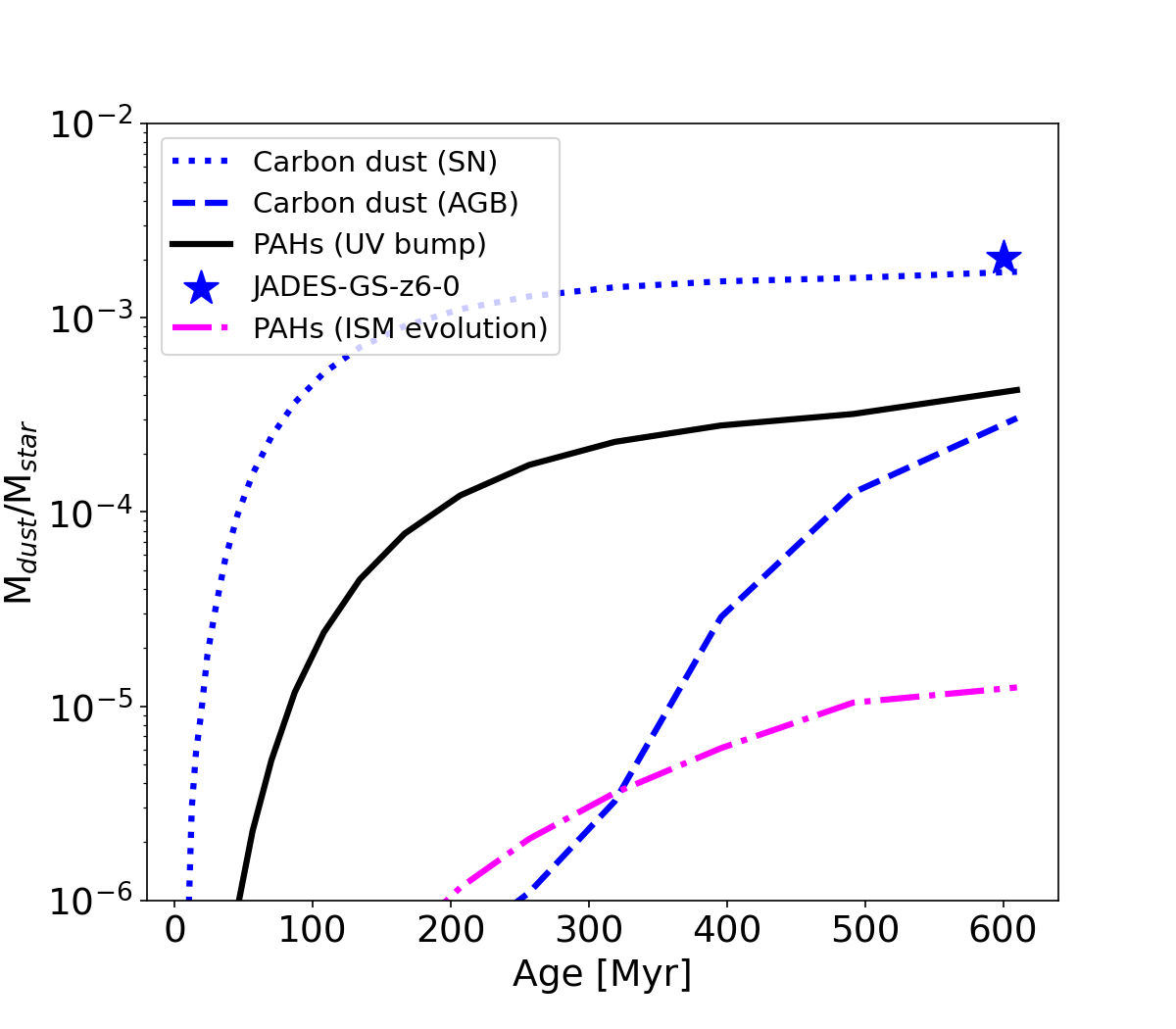}
     \vspace*{\fill}%
   \caption{Left panel: Time evolution normalised mass of the PAHs needed to reproduce the UV bump (black line) for the same scenario of dust evolution as in Fig.~\ref{Fig:spectra}, left panel, compared with the total carbon dust produced by SNe (mostly Type II; dotted line), AGB stars (dashed line) and dust growth (solid line). The amount of dust is reduced by destruction by SNe (see text). The magenta lines represent the evolution of PAHs in the ISM which does not include growth and injection from stellar sources. The star symbol represents the total carbon dust mass over stellar mass at the end point of the simulation which provides the best fit for JADES-GS-z6-0. \endgraf
   Right panel: same as in left panel but for the case without growth in the ISM (see Fig.~\ref{Fig:spectra}, right panel).}
              \label{Fig:PAH_AGB}
   \end{figure}

In Fig.~\ref{Fig:PAH_AGB}, we show, for each of the scenarios yielding the spectra in Fig.~\ref{Fig:spectra}, the contribution to the total carbon production (graphite plus PAHs) from Type~II~SNe, AGB stars, and dust growth in the ISM (when included), along with the predicted PAH mass evolution. We find a negligible amount of carbon dust from Type~Ia~SNe.
When dust growth in the ISM is included in the calculations, the bulk of the carbon dust production ($\sim 85\%$) is due to dust growth, with AGB stars contributing $\sim 8\%$ and Type~II~SNe providing the remaining $\sim 7\%$. 
To match the required abundance of PAHs in the spectrum resulting from the ISM dust growth scenario, approximately 67\% of the total carbon dust produced by SNe and AGB stars must be injected as PAHs. In this scenario, AGB stars or Type II SNe alone do not produce enough carbon to explain the observed UV bump, as shown in Fig.~\ref{Fig:PAH_AGB}, left panel. By contrast, only $\sim 7\%$ of the dust formed via growth in the ISM needs to be in the form of PAHs. In the case without dust growth in the ISM, the amount of carbon produced by AGB stars alone is not sufficient to explain the UV bump, as shown in Fig.~\ref{Fig:PAH_AGB}, right panel, while the amount required is $\sim 20$\% from Type II SNe alone.

We explored the possibility that PAHs can be produced by shattering of larger grains (injected by stellar sources or built-up in the ISM). For this study, we employed the prescriptions of \citet{Seok14} for the grain size distribution from \citet{WD01}, to which we refer for all the details. In brief, we computed the evolution of PAHs in the ISM, including their destruction by SNe shocks, astration and coagulation, and their formation through shattering of carbonaceous grains. The results are shown in Fig.~\ref{Fig:PAH_AGB}, where there is a clear mismatch between the abundance of PAHs needed to reproduce the observations (black lines) and the abundance predicted from the evolution in the ISM (magenta lines). This finding indicates that PAHs are unlikely to form through shattering of larger grains in our scenarios, but that some contribution to their abundance coming from stellar sources and/or build-up in the ISM is needed.
Furthermore, the large PAH fraction found to reproduce the observations (4--4.6\%) makes  JADES-GS-z6-0 a strong candidate for further studies exploring how substantial PAH mass survives hard radiation field. In the local Universe such cases have been found only amongst few galaxies exhibiting high ratio of large neutral PAH molecules vs. small ones \citep[e.g.][]{Bernete22}

\subsection{Rest-UV slope evolution}
\begin{figure}
   \centering
   \vspace*{\fill}%
      \includegraphics[scale=1]{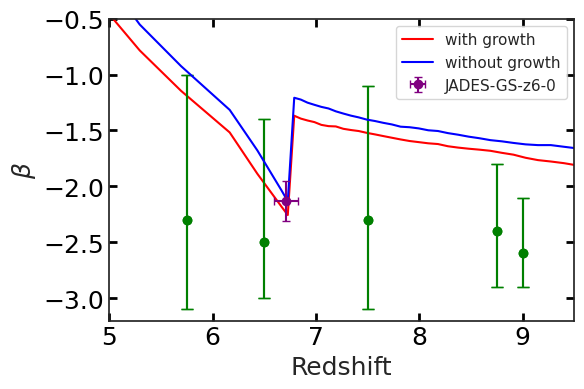}
     \vspace*{\fill}%
   \caption{Redshift evolution of galaxy UV slope ($\beta$) estimated for distant galaxies observed with  JWST. The green symbols represent median values of a sample from \citet{Saxena24}; error bars correspond to the observed scatter in the data. Over-plotted are predictions from our two best models that reproduce the $\beta$ slope at the redshift of JADES-GS-z6-0. Modeled tracks with/without ISM growth included are showed with the red and blue lines, respectively.}
              \label{Fig:beta}
   \end{figure}

Fig.~\ref{Fig:beta} presents the redshift evolution of the rest-UV continuum slope, $\beta$, highlighting both observations and the evolution of the best-fitting models for JADES-GS-z6-0. The best-fitting models are tuned to reproduce its $\beta$ slope by including a burst in the SFH producing $\sim 30\%$ of the total stellar mass. The green symbols with error bars showing the observed mean  values of $\beta$ and their associated scatter are taken from \citet{Saxena24}. Both models tend to overpredict $\beta$ at redshift $\approx 9$ possibly indicating complex SFHs for different galaxies, including multiple bursts, that would decrease the value of $\beta$ \citep{Narayanan24}. The best-fitting model with dust growth predicts a bluer slope than the scenario without growth. The large scatter in the data reflects different dust content in the ISM of galaxies and different SFHs.

\section{Conclusions}
In this work, we investigated the metal and dust enrichment in the $z\sim 6.7$ galaxy \mbox{JADES-GS-z6-0} and modeled its spectrum by means of radiative transfer calculations. If dust growth occurs in the ISM, the continuum and UV bump are well reproduced with approximately 4\% of PAHs (as a mass fraction of the total dust mass) when assuming a depletion fraction of $f_{\rm SN} \sim 10\%$ for SNe. In the absence of such growth, the required dust production efficiency from Type II SNe increases to $f_{\rm SN} \sim 73\%$ with PAH mass fraction of $\sim 4.6$ \%. 

We explored two potential channels for PAH formation in this early galaxy: (1) synthesis in the circumstellar envelopes of AGB stars, and (2) the shattering of large grains in the ISM. In our model, we find that neither process alone can produce enough PAHs. Specifically, the cumulative carbon dust yields from standard AGB production remain below the observed PAH abundance, while the shattering process does not produce a sufficient mass of PAHs. Hence, our findings suggest that these processes alone are insufficient to account for the observed PAH abundance under the assumed conditions. Consequently, additional PAH formation in the ISM and/or contributions from massive stars (i.e. winds of carbon-rich Wolf Rayet stars and/or Type II SNe) are likely necessary.

Furthermore, we examine the evolution of the UV slope $\beta$ with redshift. Our comparison underscores the critical role of ISM dust processes in shaping $\beta$, while also highlighting the need for a more comprehensive treatment of dust production and destruction, and of the diversity in SFH.

% WARNING
%-------------------------------------------------------------------
% Please note that we have included the references to the file aa.dem in
% order to compile it, but we ask you to:
%
% - use BibTeX with the regular commands:
%   \bibliographystyle{aa} % style aa.bst
%   \bibliography{Yourfile} % your references Yourfile.bib
%
% - join the .bib files when you upload your source files
%-------------------------------------------------------------------

%%%%%%%%%%%%%%%%%%%%%%%%%%%%%%%%%%%%%%%%%%%%%%%%%%%%
%%                 REFERENCES
%%%%%%%%%%%%%%%%%%%%%%%%%%%%%%%%%%%%%%%%%%%%%%%%%%%%
\bibliographystyle{aasjournal}  % Or any style you prefer
\bibliography{nanni}            % Adjust to your .bib file
\begin{acknowledgements}
A.N, M.R., P.S. acknowledge support from the Narodowe Centrum Nauki (NCN), Poland, through the SONATA BIS grant UMO-2020/38/E/ST9/00077. 
D.D. acknowledges support from the NCN through the SONATA grant UMO-2020/39/D/ST9/00720. D.D thanks the support from the Polish National Agency for
Academic Exchange (Bekker grant BPN /BEK/2024/1/00029/DEC/1).
M.R. acknowledges support from the Foundation for Polish Science (FNP) under the program START 063.2023.
J.W. gratefully acknowledges support from the Cosmic Dawn Center through the DAWN Fellowship. The Cosmic Dawn Center (DAWN) is funded by the Danish National Research Foundation under grant No. 140.
This research made use of Astropy,\footnote{\url{http://www.astropy.org}} a community-developed core Python package for Astronomy \citep{astropy2022}.
We thank the anonymous referee for the careful reading of the manuscript and for her/his thoughtful comments.
\end{acknowledgements}

\end{document}